\documentclass{epl}
\usepackage{color}

\begin{document}

\title{Covalent bonding and hybridization effects in the corundum-type 
       transition-metal oxides $ {\rm \bf V_2O_3} $ and 
       $ {\rm \bf Ti_2O_3} $}

\shorttitle{Covalent bonding and hybridization effects \ldots}

\author{V.\ Eyert 
        \thanks{E-mail: \email{volker.eyert@physik.uni-augsburg.de}}
        \and U.\ Schwingenschl\"ogl
        \and U.\ Eckern}
\shortauthor{V.\ Eyert \etal}

\institute{Institut f\"ur Physik, Universit\"at Augsburg, 
           86135 Augsburg, Germany}


\maketitle

\begin{abstract}
The electronic structure of the corundum-type transition-metal oxides
$ {\rm V_2O_3} $ and $ {\rm Ti_2O_3} $ is studied by means of the
augmented spherical wave method, based on density-functional theory 
and the local density approximation.
Comparing the results for the vanadate and the titanate allows us to
understand the peculiar shape of the metal $ 3d $ $ a_{1g} $ density of 
states, which is present in both compounds. The $ a_{1g} $ states are 
subject to pronounced bonding-antibonding splitting due to metal-metal 
overlap along the $ c $-axis of the corundum structure. However, the 
corresponding partial density of states is strongly asymmetric with 
considerably more weight on the high energy branch. We argue that 
this asymmetry is due to an unexpected broadening of the bonding 
$ a_{1g} $ states, which is caused by hybridization 
with the $ e_g^{\pi} $ bands. In contrast, the antibonding $ a_{1g} $ 
states display no such hybridization and form a sharp peak. Our results 
shed new light on the role of the $ a_{1g} $ orbitals for the 
metal-insulator transitions of $ {\rm V_2O_3} $. In particular, due to 
$ a_{1g} $-$ e_g^{\pi} $ hybridization, an interpretation in terms of 
molecular orbital singlet states on the metal-metal pairs along the 
$ c $-axis is not an adequate description.  
\end{abstract}

Based on the electronic level scheme of Castellani {\em et al.} 
\cite{castellani78}, $ {\rm V_2O_3} $ was studied extensively as the 
canonical Mott--Hubbard system. Due to octahedral coordination of the 
metal sites, the V $ 3d $ states are split into lower $ t_{2g} $ and 
higher $ e_g^{\sigma} $ levels. Trigonal lattice symmetry leads to 
further separation of the former into $ a_{1g} $ and $ e_g^{\pi} $ 
states. Covalent V-V bonding along the hexagonal $ c $-axis of the 
corundum structure results in bonding and antibonding $ a_{1g} $ 
molecular orbitals, which sort of bracket the $ e_g^{\pi} $ levels. 
Electronic structure calculations confirmed the gross features of this 
scheme with $ t_{2g} $ states close to the Fermi level \cite{mattheiss94}. 
As proposed by Castellani {\em et al.}, the bonding $ a_{1g} $ states 
are fully occupied, whereas the antibonding states shift to higher 
energies. This leaves one electron per V atom in the twofold degenerate 
$ e_g^{\pi} $ orbital and leads to an $ S=1/2 $ state, suggesting to 
use the one-band Hubbard model at half filling for describing the 
metal-insulator transitions (MITs) of $ {\rm V_2O_3} $. The 
stoichiometric compound undergoes an MIT at 168\,K, leading from a 
paramagnetic metallic (PM) to an antiferromagnetic insulating (AFI) 
phase; on doping with Cr or Al a paramagnetic insulating (PI) phase 
is found \cite{mcwhan71,mcwhan73}.

The model by Castellani {\em et al.}\ has been called into question by 
polarized x-ray absorption spectroscopy experiments, which point at an 
$ S=1 $ spin state of the metal atoms \cite{park00}.  The lowest 
excited states miss pure $ e_g^{\pi} $ symmetry but reveal remarkable 
$ a_{1g} $ admixtures calling for a treatment beyond the one-band 
Hubbard model. LDA+U calculations explain the peculiar antiferromagnetic 
order at low temperatures and point at an $ S=1 $ configuration 
\cite{ezhov99}, which also results from the model calculations by
Mila {\it et al.} \cite{mila00}. Starting with the assumption of strong 
covalent bonding in the V-V pairs parallel to the $ c $-axis, these authors 
suppose the intersite $ a_{1g} $ hopping matrix element to dominate. 
However, a recent study of the hopping processes in $ {\rm V_2O_3} $ 
found relevant matrix elements also between second, third, and fourth 
nearest neighbours \cite{elfimov03}. LDA band structures show only 
minor response to the structural modifications occuring at the phase 
transitions of $ {\rm V_2O_3} $, whereas a correct description of the 
PM-PI transition has been obtained by a combination of LDA 
calculations with the dynamical mean field theory (DMFT) 
\cite{held01,keller04}. Nonetheless, confirming an early suggestion 
by Dernier \cite{dernier70} we have recently demonstrated 
that considering the rearrangements of the metal-metal overlap 
{\em perpendicular} to the hexagonal $ c $-axis is important for 
understanding the MITs of $ {\rm V_2O_3 } $ \cite{us03,us03b,us04}.  

$ {\rm Ti_2O_3 } $ is isostructural to corundum $ {\rm V_2O_3} $ and 
undergoes a gradual MIT without symmetry breaking lattice distortion 
between 400\,K and 600\,K. In contrast to $ {\rm V_2O_3} $, its low 
temperature insulating state is non-magnetic. As in $ {\rm V_2O_3} $, 
the $ a_{1g} $ states mediate strong bonding between Ti-Ti pairs in 
face-sharing octahedra parallel to the hexagonal $ c $-axis, leading 
to bonding and antibonding $ a_{1g} $ bands bracketing the $e_g^{\pi}$
states. An insulating energy gap is expected between the bonding 
$ a_{1g} $ and $ e_g^{\pi} $ states \cite{vanzandt68}. According to 
this picture increase of the $ c/a $ ratio of the corundum lattice 
constants with increasing temperature reduces the $ a_{1g} $ band 
splitting and promotes a collapse of the energy gap. Band structure 
calculations by Mattheiss \cite{mattheiss96} revealed a partially filled 
$ t_{2g} $ complex of overlapping $ a_{1g} $ and $ e_g^{\pi} $ bands 
at the Fermi level. Decreasing the $ c/a $ ratio reduces but does not 
eliminate the $ a_{1g} $-$ e_g^{\pi} $ band overlap. To open the gap,  
an unphysically small Ti-Ti distance of 2.2\,\AA\ parallel to the 
$ c $-axis is required, which precludes a simple band explanation of 
the MIT. Recent cluster LDA+DMFT calculations assuming moderate Coulomb 
interactions among the $ t_{2g} $ orbitals reproduced the insulating 
state \cite{poteryaev03}.

In this Letter we report on electronic structure calculations for 
both $ {\rm V_2O_3 } $ and $ {\rm Ti_2O_3 } $ using the respective 
room-temperature crystal structure data. Our calculations i) result 
in a new interpretation of the bonding and antibonding $ a_{1g} $ 
states and ii) reveal striking differences in the hybridizations 
of the V $ 3d $ $ e_g^{\pi} $ bands with the bonding and antibonding 
$ a_{1g} $ states, respectively. While hybridization of the 
$ e_g^{\pi} $ bands with the bonding $ a_{1g} $ states is rather strong, 
leading to significant broadening of these bands, the antibonding 
states are of pure $ a_{1g} $ character. 

The present LDA band structure calculations are based on the 
scalar-relativistic augmented spherical wave (ASW) method 
\cite{williams79,eyert00b}. Crystallographic data reported by 
Dernier for $ {\rm V_2O_3} $ \cite{dernier70} and by Rice and 
Robinson for $ {\rm Ti_2O_3 } $ \cite{rice77} were used. To model 
the crystal potential in the large voids of the open crystal 
structures, additional augmentation spheres were included. Optimal 
augmentation sphere positions and radii of all spheres were 
automatically generated by the sphere geometry optimization (SGO) 
algorithm \cite{eyert98b}. The basis sets comprised V/Ti 
$ 4s $, $ 4p $, $ 3d $, and O $ 2s $, $ 2p $ orbitals as well as 
states of the additional augmentation spheres. Brillouin zone 
integrations were performed using up to 2480 $ {\bf k} $-points 
within the irreducible wedge.

\begin{figure}
\oneimage[width=140mm]{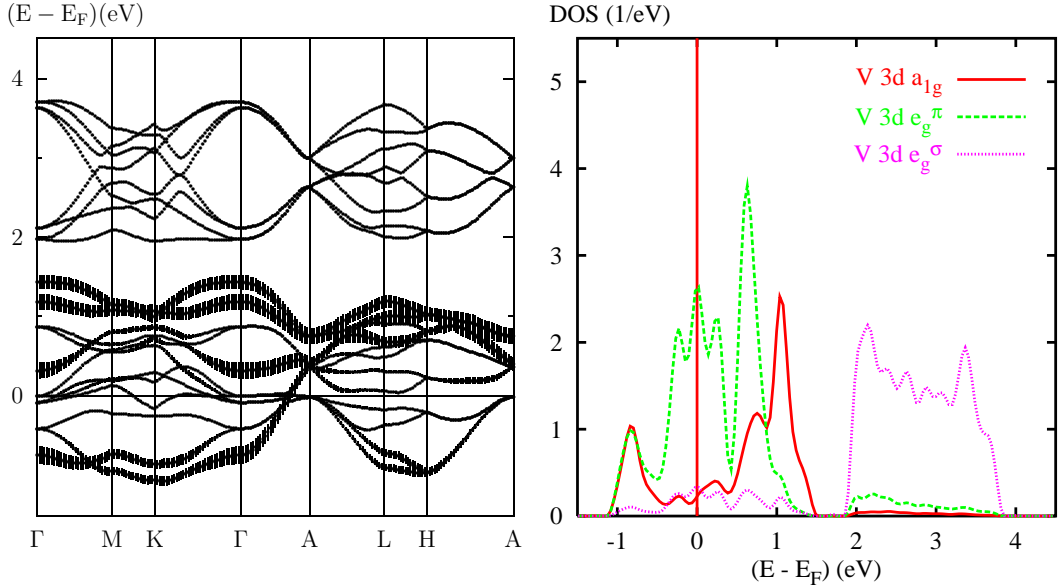}
\caption{$ {\rm V_2O_3} $: weighted electronic bands and partial 
         V $ 3d $ DOS. The length of the bars in the band structure 
         indicates V $ 3d $ $ a_{1g} $ contributions.}
\label{fig1}
\end{figure}

As is typical for transition-metal chalcogenides with octahedral 
coordination, $ {\rm V_2O_3} $ shows three groups of bands in the 
vicinity of the Fermi level, extending from $ -8.8 $\,eV to $ -3.9 $\,eV, 
from $ -1.1 $\,eV to 1.5\,eV, and from 1.9\,eV to 3.8\,eV, see Fig.\ 
\ref{fig1}.  
They originate from O $ 2p $, V $ 3d $ $ t_{2g} $, and V $ 3d $ 
$ e_g^{\sigma} $ states, respectively. The right hand side of 
Fig.\ \ref{fig1} shows the V $ 3d $ density of states (DOS) for 
the latter two groups of bands separated into its symmetry 
components $ a_{1g} $, $ e_g^{\pi} $, and $ e_g^{\sigma} $. 
V-O hybridization results in additional O $ 2p $ admixtures in 
the energy range shown, which are stronger for the $ \sigma $-bonding 
$ e_g^{\sigma} $ states. Octahedral coordination of the metal atoms 
leads to almost perfect energetical separation of the $ t_{2g} $ and 
$ e_g^{\sigma} $ groups of bands. Small contributions of $ t_{2g} $ states 
in the $ e_g^{\sigma} $ energy range above 1.9\,eV are mainly due to 
the V-V anti-dimerization along the $ c $-axis, which shifts the metal 
atoms off the centers of their oxygen octahedra and results in a 
metal-metal distance much larger than the ideal value of 1/6 of the 
$ c $ lattice constant. 

The band structure of the $ t_{2g} $ and $ e_g^{\sigma} $ groups of 
bands is displayed on the left hand side of Fig.\ \ref{fig1}, where 
the length of the bars added to the bands is proportional to the 
respective $ a_{1g} $ contribution. Hence, bands without bars are 
of $ e_g^{\pi} $ and $ e_g^{\sigma} $ character, respectively, in 
the lower and upper group shown. The band structure refers to the 
non-primitive hexagonal representation of the trigonal unit cell. 
While the dispersion of the $ e_g^{\pi} $ bands is rather isotropic, 
the $ a_{1g} $ bands have an increased dispersion along the line 
$ \Gamma $-A. Perpendicular to this line, i.e.\ along the paths 
$ \Gamma $-M-K-$ \Gamma $ and A-L-H-A, the $ a_{1g} $ states are 
found mainly at the lower and upper boarder of the band complex at 
the Fermi level. This holds especially for the bands above 1\,eV 
in the $ \Gamma $-M-K-$ \Gamma $ plane, which are well separated 
from the lower lying bands. 

The pronounced one-dimensionality of the $ a_{1g} $ bands leads to the 
characteristic shape of the $ a_{1g} $ partial DOS. It suggests to 
interprete the peak at $ -0.9 $\,eV and the double-peak at 0.8/1.0\,eV, 
respectively, as the bonding and antibonding states resulting from the 
V-V overlap across the octahedral faces. However, the weights of these 
peaks are far from equal but have a ratio of about 1:3, contradicting 
expectations based on a molecular orbital point of view. Due to the 
reduced weight of the lower $ a_{1g} $ peak the occupation of the 
twofold degenerate $ e_g^{\pi} $ states is larger than one, explaining 
the experimental findings of Park {\em et al.}\ \cite{park00}. 

\begin{figure}
\oneimage[width=140mm]{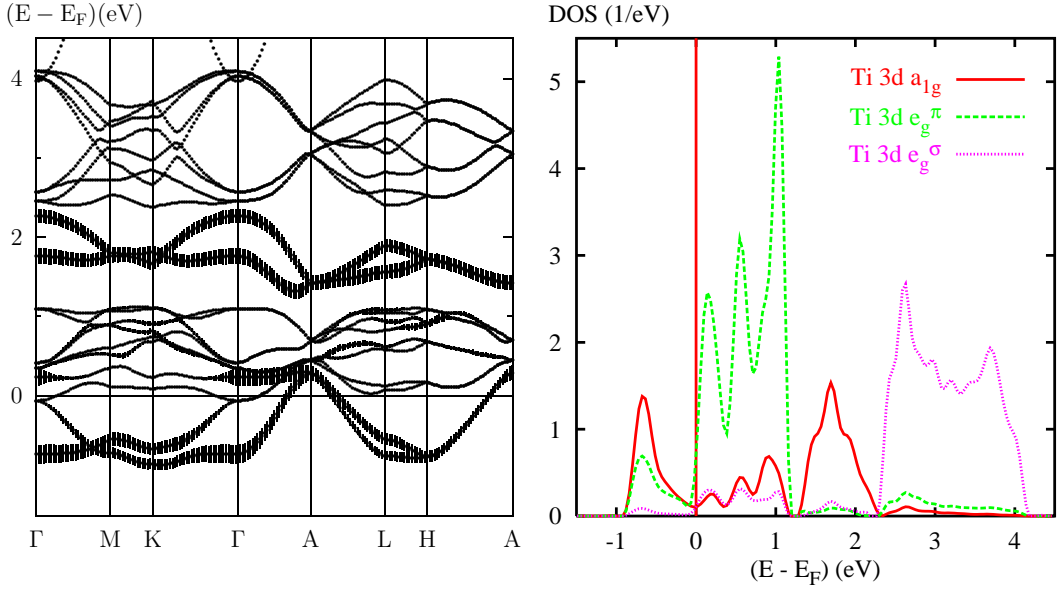}
\caption{$ {\rm Ti_2O_3 } $: weighted electronic bands and partial 
         Ti $ 3d $ DOS. The length of the bars in the band structure 
         indicates Ti $ 3d $ $ a_{1g} $ contributions.}
\label{fig2}
\end{figure}

In order to understand these puzzling results we turn to titanium 
sesquioxide, which has the same octahedral coordination of the 
transition-metal sites with octahedra neighbouring along the $ c $-axis 
interlinked via faces. The band structure and the partial densities of 
states of $ {\rm Ti_2O_3 } $ are displayed in Fig.\ \ref{fig2},
in the representation already used for $ {\rm V_2O_3} $ in Fig.\ 
\ref{fig1}. From the structural similarity of the two compounds we 
expect the same gross features of their electronic structures. 
Indeed, we obtain for $ {\rm Ti_2O_3 } $ three groups of bands in the 
vicinity of the Fermi level, extending from $ -9.0 $\,eV to $ -4.5 $\,eV 
(O $ 2p $), $ -0.9 $\,eV to 2.3\,eV (Ti $ 3d $ $ t_{2g} $), and 
2.4\,eV to 4.1\,eV (Ti $ 3d $ $ e_g^{\sigma} $). Additional Ti $ 4s $ 
bands are found at energies higher than 3.9\,eV. These results agree 
well with the findings of Mattheiss \cite{mattheiss96}. 

The symmetry components of the Ti $3d$ DOS (right hand side of 
Fig.\ \ref{fig2}) confirm the energetical separation of the $ t_{2g} $ 
and $ e_g^{\sigma} $ states as well as the $ t_{2g} $ admixture to the 
bands above 2.4\,eV due to the metal-metal anti-dimerization along the 
$ c $-axis.  
In the band structure shown on the left hand side of Fig.\ \ref{fig2} 
we distinguish the isotropically dispersing $ e_g^{\pi} $ bands from 
the $ a_{1g} $ states, which are highlighted by the bars attached to 
each band and display a rather one-dimensional behaviour. The latter 
causes the characteristic shape of the $ a_{1g} $ partial DOS with 
the pronounced peaks at $ -0.7 $\,eV and 1.7\,eV, which as before might 
be attributed to bonding-antibonding splitting due to metal-metal 
overlap along the $ c $-axis. Yet, we are again faced with a decreased 
weight of the low-energy peak leading to the finite $ e_g^{\pi} $ 
occupation. 

While the results for $ {\rm V_2O_3} $ and $ {\rm Ti_2O_3} $ are 
quite similar in general, we observe distinct differences at a 
closer look. They might give a first clue to the problem of the 
unequal weights of the $ a_{1g} $ peaks and pave the way for a 
better understanding of the role of these electronic states. To 
be specific, we point to the small gap at 1.2\,eV in the partial 
DOS of the titanate, separating the peak at 1.7\,eV from the lower 
lying bands. Such a gap is not observed for the vanadate. Since 
the $ a_{1g} $ partial densities of states both below and above 
1.2\,eV integrate to about one electron each, it is tempting to 
interpret the bands above this gap as antibonding and the whole of 
the $ a_{1g} $ bands between $ -1 $\,eV and 1.2\,eV as bonding. If 
this point of view were correct, the bonding $ a_{1g} $ states would 
thus extend over a very broad energy interval of more than 2\,eV. 

This quite unusual situation should be related to the second 
observation growing out of the comparison of Figs.\ \ref{fig1} and 
\ref{fig2}, namely the conspicuous downshift of the upper edge of 
the $ e_g^{\pi} $ states. While in the vanadate these bands share 
edges with the whole of the $ a_{1g} $ states, there exists only a 
tiny contribution at energies above 1.2\,eV in the titanate. In 
contrast, the $ e_g^{\pi} $ partial DOS displays a very sharp cutoff 
at this energy. Obviously, the $ e_g^{\pi} $ states are found only 
in the energy region of the bonding but not of the antibonding 
$ a_{1g} $ states. Going one step further, we infer from the similar 
shape of both partial densities of states below 1.2\,eV, in particular 
from the common peaks near $ - 0.7 $\,eV, 0.2\,eV, 0.5\,eV, and 1.0\,eV,  
a considerable amount of $ a_{1g} $-$ e_g^{\pi} $ mixing. We thus 
attribute the large band width of the low-energy $ a_{1g} $ states to 
the strong $ a_{1g} $-$ e_g^{\pi} $ hybridization. Since the broadening 
is of similar size as the bonding-antibonding splitting of the $ a_{1g} $ 
band, we end up with the complex situation observed for the vanadate. 

In order to ``prove'' the just sketched scenario, we turn in the last 
step to another set of calculations for $ {\rm V_2O_3} $ with a 
hypothetical crystal structure. In this structure the V-V bond length 
along the $ c $-axis has been reduced from 2.70\,\AA\ to 2.51\,\AA, 
while the symmetry of the corundum structure, the lattice constants,  
and the oxygen positions were retained. As a result of these changes, 
hypothetical 
$ {\rm V_2O_3} $ is characterized by a short intrapair V-V distance, 
hence, by increased intrapair bonding. Eventually, this will cause a 
stronger bonding-antibonding splitting of the $ a_{1g} $ states. 
As a consequence, we expect a much clearer separation of these states 
as compared to the results given in Fig.\ \ref{fig1}. The band shifts 
occurring on going from the real to the artificial structure of 
$ {\rm V_2O_3} $ will thus allow us to make a clear distinction between 
bonding and antibonding $ a_{1g} $ states for this material. 

\begin{figure}
\oneimage[width=140mm]{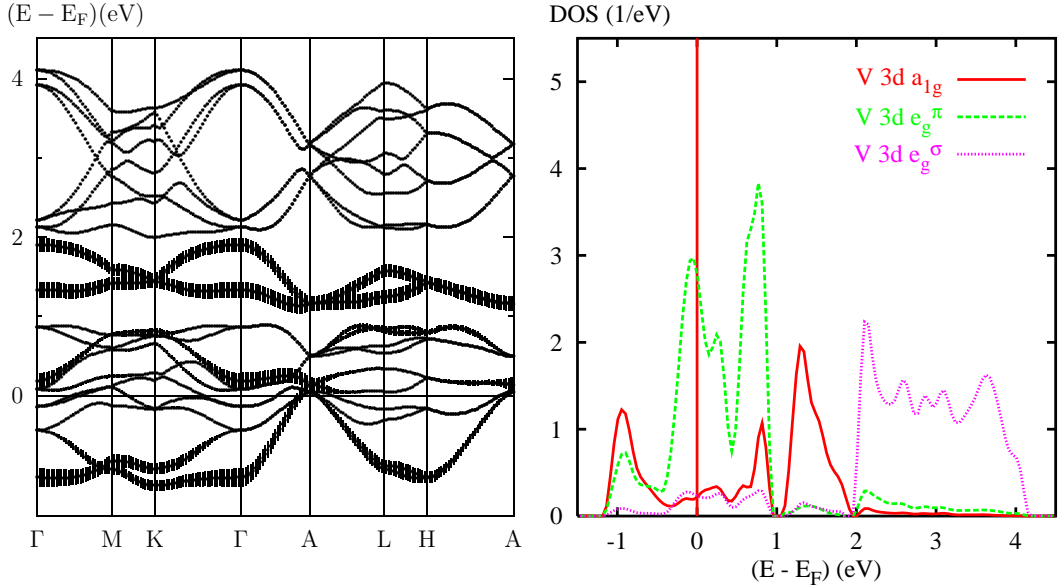}
\caption{Hypothetical $ {\rm V_2O_3} $ (V-V bond lengths along 
         $ c_{\rm hex} $ compressed to 2.51\,\AA): weighted 
         electronic bands and partial V $ 3d $ DOS. The length of 
         the bars in the band structure indicates V $ 3d $ $ a_{1g} $ 
         contributions.}
\label{fig3}
\end{figure}

The calculated results for the hypothetical $ {\rm V_2O_3} $ structure 
are displayed in Fig.\ \ref{fig3},  
using the same representation of the band structure and V $ 3d $ DOS 
as in Figs.\ \ref{fig1} and \ref{fig2}. While the V $ 3d $ $ e_g^{\sigma} $ 
DOS of artificial $ {\rm V_2O_3} $ still resembles the results for real 
$ {\rm V_2O_3} $, we observe indeed distinct modifications especially for 
the $ a_{1g} $ DOS. The peak at $ -0.8 $\,eV in Fig.\ \ref{fig1} shifts to 
lower energies, and the peak at 1.0\,eV (having a pronounced shoulder 
at 0.7\,eV, see Fig.\ \ref{fig1}) splits into two peaks at 0.8\,eV and 
1.3\,eV leaving a gap at about 1.0\,eV. At the same time the upper 
edge of the $ e_g^{\pi} $ states experiences a strong downshift and 
displays a sharp cutoff. Eventually, the band structures 
and densities of states of artificial $ {\rm V_2O_3} $ become more 
similar to those of $ {\rm Ti_2O_3 } $. From the differences between 
real and artificial $ {\rm V_2O_3} $ we thus associate the $ a_{1g} $ 
peak around 1.3\,eV with the antibonding states due to intrapair bonding. 
At lower energies, strong $ a_{1g} $-$ e_g^{\pi} $ hybridization leads to a 
considerable broadening of the bonding $ a_{1g} $ branch, which extends 
from $ -1.2 $\,eV to 1.0\,eV. Note that both the bonding and antibonding 
states integrate to equal weights. Transferring these findings to real  
$ {\rm V_2O_3} $, we conclude that the $ a_{1g} $ peak at 1.0\,eV in 
Fig.\ \ref{fig1} represents the antibonding states, whereas the region 
of bonding states extends from $ -1.1 $\,eV to $ \approx 0.9 $\,eV and 
even includes the shoulder at 0.7\,eV. 

In general, our findings agree well with the results of the recent LDA+U 
calculations by Ezhov {\em et al.}\ as well as Elfimov {\em et al.}\ 
for antiferromagnetic and assumed ferromagnetic $ {\rm V_2O_3} $  
\cite{ezhov99,elfimov03}. The main difference between both approaches 
concerns the relative downshift of the $ e_g^{\pi} $ states, which 
eventually opens the insulating gap observed for the AFI phase. However, 
the LDA+U calculations likewise result in a finite $ a_{1g} $-$ e_g^{\pi} $ 
hybridization for the bonding $ a_{1g} $ states whereas the $ e_g^{\pi} $ 
admixture to the antibonding $ a_{1g} $ states vanishes. Hence, while the 
LDA+U treatment leads to a somewhat changed scenario the basic mechanisms 
are still well described by the LDA calculations. 

Still, it remains an open question why the bonding $ a_{1g} $ states 
are subject to hybridization with the $ e_g^{\pi} $ 
states, while the antibonding states retain their pure band character. 
Of course, it is tempting to relate this finding to the aforementioned 
metal-metal anti-dimerization, which shifts the metal atoms along the 
$ c $-axis off the center of their respective oxygen octahedra. Since 
the displacement is parallel to the principal axis of the $ a_{1g} $ 
orbitals, the resulting increase in $ d $-$ p $ overlap mainly affects 
these states and is much less pronounced for the $ e_g^{\pi} $ bands. 
The $ a_{1g} $ states are thus found at slightly elevated energy, the 
difference of the centers of gravity amounting to $ \approx 0.3 $\,eV 
for $ {\rm V_2O_3} $ \cite{keller04}. As a consequence, the lower 
bonding part of the $ a_{1g} $ bands experiences more overlap with the 
$ e_g^{\pi} $ states than the high-energy antibonding $ a_{1g} $ states. 

Nevertheless, a more convincing argument is based on the symmetry of 
the electronic orbitals involved, as well as the much different degree 
of overlap of the $ t_{2g} $ orbitals within the $ c $-axis pairs. 
As a matter of fact, the $ e_g^{\pi} $ orbitals arise as linear 
combinations of all five $ d $ states, except for the $ d_{3z^2-r^2} $ 
states which are identical to the $ a_{1g} $ states. In addition, 
the overlap of the $ e_g^{\pi} $ orbitals within the vanadium pairs 
along the $ c $-axis is close to negligible. As a consequence, these 
states are symmetric with respect to reflection about the midplane 
between the two atoms. This is different for the $ a_{1g} $ states, 
which do overlap along the $ c $-axis this leading to the bonding and 
antibonding combinations. Since the bonding and antibonding $ a_{1g} $ 
states are even and odd functions with respect to reflections, only 
the former may hybridize with the $ e_g^{\pi} $ orbitals, whereas the 
latter do not overlap with these states. 

In conclusion, the electronic structure of the corundum-type sesquioxides 
$ {\rm V_2O_3} $ and $ {\rm Ti_2O_3} $ is strongly influenced by a complex 
interplay of i) the formation of bonding and antibonding states arising 
from overlap of the metal $ 3d $ $ a_{1g} $ states across octahedral faces 
and ii) strong hybridization of the bonding but not the antibonding 
$ a_{1g} $ bands with the $ e_g^{\pi} $ states. The hybridization disturbs 
the formation of molecular orbital singlet states by the $ a_{1g} $ orbitals 
of the vertical V-V pairs. Therefore, the occupation of the $ e_g^{\pi} $ 
bands is increased this explaining the recently observed $ S=1 $ 
configuration. As a consequence, failure of the model by Castellani 
{\em et al.}\ is mainly due to hybridization effects and not purely a 
consequence of the energy lowering of the $ e_g^{\pi} $ states. In 
addition, since for symmetry reasons the $ a_{1g} $-$ e_g^{\pi} $ 
hybridization affects only the bonding $ a_{1g} $ states, the 
corresponding partial DOS assumes a strongly asymmetric shape. The 
observed strong 3D-coupling of the $ a_{1g} $ orbitals appears to be 
a primal feature of the electronic structure of these sesquioxides and 
therefore is of special importance for models dealing with the MITs 
in this compound. The findings for $ {\rm V_2O_3} $ strongly differ from 
those for $ {\rm VO_2} $ and $ {\rm Ti_4O_7} $, where hybridization 
between the $ d_{\parallel} $ states, which are responsible for 
metal-metal overlap, and the $ e_g^{\pi} $ states is suppressed and 
thus allows for a Peierls-type mechanism of the MIT \cite{eyert02,eyert04}. 

\begin{acknowledgments}
This work was supported by the Deutsche Forschungsgemeinschaft (DFG) 
through Sonderforschungsbereich 484. 
\end{acknowledgments}

\end{document}